\documentclass[10pt,twocolumn,letterpaper]{article}

\usepackage{wacv}              

\usepackage{graphicx}
\usepackage{amsmath}
\usepackage{amssymb}
\usepackage{booktabs}
\usepackage{xcolor}
\usepackage{soul}

\usepackage[pagebackref,breaklinks,colorlinks]{hyperref}

\usepackage[capitalize]{cleveref}
\crefname{section}{Sec.}{Secs.}
\Crefname{section}{Section}{Sections}
\Crefname{table}{Table}{Tables}
\crefname{table}{Tab.}{Tabs.}

\def\wacvPaperID{9} 
\def\confName{WACV}
\def\confYear{2025}

\begin{document}

\title{FlatTrack: Eye-tracking with ultra-thin lensless cameras}
\author{Purvam Jain\\
IIT Madras\\
{\tt\small purvam@smail.iitm.ac.in}
\and
Althaf M. Nazar\\
IIT Madras\\
{\tt\small althafmnazar123@gmail.com}
\and
Salman S. Khan\\
Rice University\\
{\tt\small skhan@rice.edu}
\and
Kaushik Mitra\\
IIT Madras\\
{\tt\small kmitra@iitm.ac.in}
\and
Praneeth Chakravarthula\\
UNC Chapel Hill\\
{\tt\small cpk@cs.unc.edu}
}

\twocolumn[{%
\renewcommand\twocolumn[1][]{#1}%
\maketitle
\begin{center}
    \centering
    \captionsetup{type=figure}
    \includegraphics[width=1\textwidth,height=5cm]{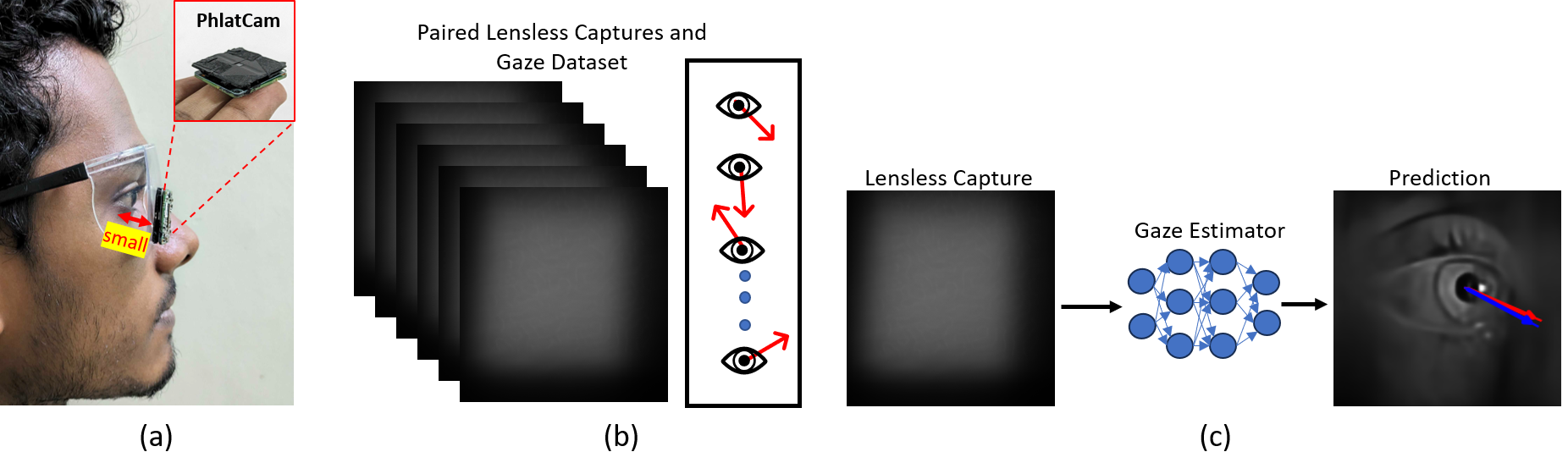}
    \captionof{figure}{%
  	\textbf{FlatTrack Gaze Estimation.} (a) PhlatCam lensless camera allows development of small form-factor gaze tracker. (b) We propose a large dataset of paired lensless captures and gaze directions, which we use to evaluate the efficacy of (c) various lensless gaze estimation techniques.%
    }
\end{center}%
}]

\begin{abstract}

\vspace{-10pt}

Existing eye trackers use cameras based on thick compound optical elements, necessitating the cameras to be placed at
focusing distance from the eyes. This results in the overall bulk of wearable eye trackers, especially for augmented and virtual reality
(AR/VR) headsets. We overcome this limitation by building a compact flat eye gaze tracker using mask-based lensless cameras. These
cameras, in combination with co-designed lightweight deep neural network algorithm, can be placed in extreme close proximity to the
eye, within the eyeglasses frame, resulting in ultra-flat and lightweight eye gaze tracker system. We collect a large dataset of near-eye
lensless camera measurements along with their calibrated gaze directions for training the gaze tracking network. Through real and
simulation experiments, we show that the proposed gaze tracking system performs on par with conventional lens-based trackers while
maintaining a significantly flatter and more compact form-factor. Moreover, our gaze regressor boasts real-time ($>$125 fps) performance for gaze tracking.
\end{abstract}

\section{Introduction}
Eye tracking plays a crucial role in Augmented and Virtual Reality (AR/VR) systems, whether it be augmenting the realism and interactivity of the immersive experience or rendering efficient display imagery. For example, by precisely tracking where the user is looking at any given time, eye tracking facilitates more natural and intuitive interactions, such as gaze-based interfaces for selection and control, which can enhance the user engagement with the virtual environment without the need for physical controllers \cite{7893315,zhang2017eye}. Eye tracking also facilitates optimized usage of system resources by rendering high-resolution graphics only in the regions of the visual field where the user's gaze is well-focused, and there is less detail in the periphery, a technique called foveated rendering that is shown to dramatically improve the display performance\cite{jabbireddy2022foveated,akcsit2019manufacturing,patney2016towards,10.1145/2366145.2366183}. Apart from improving the overall graphical performance through selective on-demand rendering, foveated rendering significantly reduces computational load, allowing for smoother and more immersive AR/VR experiences. 

Existing eye-tracking technologies (Tobii Pro, Pupil Labs) face several significant challenges that impact their practicality across various applications despite their rapid developments in recent years. Apart from the need for more accuracy and reliability, the physical size of eye-tracking units often limits their integration into wearable AR/VR headsets, posing significant design issues. Specifically, the optical components like lenses, filters and their housing directly affect the overall size of the device. The structure used to mount the sensors and associated electronics, including the enclosure that protects these components, further adds to the bulk of the system. Furthermore, privacy concerns also arise as eye tracking can reveal sensitive personal information. As demonstrated in earlier works, eye images can possibly be misused to reconstruct the scenes being observed, which pose significant threats and privacy concerns\cite{Alzayer_2024_CVPR}. 
Finally, eye-trackers require low latency to ensure responsive and immersive user experiences in interactive applications. 
These obstacles necessitate continuous advancements in hardware and software to enhance the versatility and user-friendliness of eye-tracking technology. Hence, we propose a lensless gaze-tracking prototype that has a small form factor, inexpensive, lightweight, and privacy-preserving\cite{khan2023opencam,boominathan2020phlatcam}.

In lensless imaging, the focusing lens is replaced with a thin, lightweight, and potentially inexpensive optical encoder, resulting in flat and lightweight miniaturized cameras. A lensless camera uses advanced computational reconstruction algorithms to recover the scene from captured sensor measurements, which no longer resemble the imaged scene \cite{boominathan2020phlatcam,Antipa_2017,7517296}. However, with a well-designed encoder, the measurements contain sufficient information to recover an image of the scene using post-processing algorithms that can demultiplex the sensor measurements to reconstruct a sharp image of the scene. While digital post-processing of images is standard for tasks such as distortion correction, synthetic depth of field, and denoising, lensless cameras are fundamentally different in that the post-processing is a part of the imaging system design where the optical imaging hardware, and the algorithmic software is designed together. Moreover, since lensless cameras encode information indirectly in the measurement, then computationally extract it by solving an inverse problem, they also provide promise for enabling privacy in the eye tracking measurements\cite{khan2023opencam}. Additionally, lensless cameras inherently capture the 3D information, which ensures a more accurate prediction\cite{bagadthey2022flatnet3d}.

In the current work, we employ a lensless camera design accompanied by a lightweight convolutional neural network to achieve compact eye tracking. Specifically, we use a Near-Infrared (NIR) PhlatCam\cite{boominathan2020phlatcam} to develop an ultra-thin near-eye tracker. Using our system, we collect the first-of-its-kind large dataset of nearly 20000 paired lensless captures and calibrated gaze vectors. We propose a two-stage approach for estimating gaze from lensless capture and evaluate the efficacy of such an approach on our proposed dataset. Our experiments show that lensless cameras allow us to build compact gaze trackers with high-fidelity gaze estimates. We plan to release the dataset and gaze tracking algorithm upon acceptance of the paper.

In summary, our contributions in this paper are:
\begin{itemize}

    \item We build a system using NIR PhlatCam prototype for gaze tracking. Existing works on lensless gaze estimation have mostly shown results on simulated data that don't capture the real data's non-idealities.
    \item We collect a large dataset of 20475 paired lensless captures and gaze directions from a near-eye setup. To the best of our knowledge, we are the first to collect such a dataset for lensless gaze tracking.  
    \item We build a two-stage gaze estimator to evaluate the efficacy of lensless gaze estimation and demonstrate that lensless cameras allow high-fidelity gaze recovery with a much smaller form factor. Our gaze regressor performs real-time ($>$125 fps) gaze estimation.
\end{itemize}

\section{Related Works}
\textbf{Mask-based Lensless Imaging.}
Mask-based lensless cameras replace the lens of conventional cameras with a thin optical mask. This mask has the ability to modulate the incoming light and can be placed extremely close to the sensor, leading to a flat form factor. Moreover, the ultra-small mask-to-sensor distance allows these cameras to have a large depth of field, thereby making it possible to image from a very close range - a property that will allow lensless camera-based gaze trackers to have a much smaller form factor. Multiple lensless cameras have been developed in the literature. FlatCam \cite{7517296} is a lensless camera that places a separable coded amplitude mask above a bare sensor array to enable a thin and flat form-factor imaging device, which can simulate a conventional camera by reconstructing conventional images from coded measurements. DiffuserCam \cite{Antipa_2017} and PhlatCam \cite{boominathan2020phlatcam} replace the coded amplitude mask from FlatCam \cite{7517296} with a coded phase mask for improved light efficiency and reconstruction quality. Spectral DiffuserCam\cite{Monakhova:20} exploits the multiplexing ability of lensless imagers to do hyper-spectral imaging. The primary differences in the above design lie in the pattern of the mask used and the mask-to-sensor distance. In this work, we will primarily focus on the PhlatCam \cite{boominathan2020phlatcam} lensless camera. Previous works like \cite{bagadthey2022flatnet3d,khan2023opencam} have shown the application of this camera design for 3D imaging and optical encryption. Previously, FlatCam \cite{7517296} was shown in \cite{You_2022} to perform at par with lensed cameras for gaze-tracking. However, the authors only reported the performance on simulated data, and ensuring the same performance on real data is hard. We take a step further and develop the first-of-its-kind real lensless gaze dataset using PhlatCam. We run extensive experiments on this real dataset to highlight the efficacy of our system. We plan to release this dataset upon acceptance of the paper.

\textbf{Gaze Tracking Datasets.} Since learning-based gaze trackers need large datasets, many such datasets have been proposed in the literature for both remote and near-eye gaze tracking. Among the remote gaze tracking datasets, MPII-Gaze\cite{zhang2015appearance} contains 214K face images captured using a laptop camera from 15 subjects along with their gaze vectors. Columbia Gaze\cite{smith2013gaze} contains 5880 images from 56 people and focused on gaze locking. GazeCapture\cite{krafka2016eye} contains 2.5M frames collected from over 1450 people. Among the near-eye gaze tracking datasets, NVGaze\cite{kim2019nvgaze} contains both real and synthetic data of 2M infrared images of
eyes at 1280 × 960 resolution. OpenEDS\cite{palmero2020openeds2020} contains more than 500K eye-images with the corresponding 3D gaze vectors collected from 80 participants. Recently, \cite{Angelopoulos_2021} proposed a dataset captured using a DAVIS event camera. This dataset contains both event and grayscale frames along with the corresponding gaze. Although the existing datasets contain a wide distribution of eye images and environment, none of them are suited for lensless cameras, which have a very distinct imaging model. Thus, there's an urgent need for a dataset developed specifically for lensless cameras.

\section{Lensless Imaging Background}
In a mask-based lensless camera\cite{7517296,boominathan2020phlatcam} like the one used in this work, the conventional lens is replaced by a thin optical mask placed at some distance from the sensor that modulates the incoming light. For a scene $X$, the measurement recorded by a sufficiently large sensor $Y$ is given by:
\begin{equation}\label{eq:lensless_eq}
    Y = P * X + N,
\end{equation}
where $*$ is the full-size convolutional operator (no cropping due to finite sensor size),  $P$ is the point spread function (PSF), and $N$ is additive noise. The PSF is the response of the camera to a point source. The optical mask can be implemented using an amplitude mask that attenuates the incoming light \cite{7517296} or a phase mask that modulates based on diffraction \cite{boominathan2020phlatcam}. In this work, we use a phase mask. The lensless PSF is a function of the mask pattern, wavelength of light, and the mask-sensor distance. Our PhlatCam uses a phase mask designed for NIR wavelength of 700nm and is placed slightly less than 1.5mm from the sensor. 

\begin{figure}[!tb]
  \centering
  \includegraphics[width=\columnwidth]{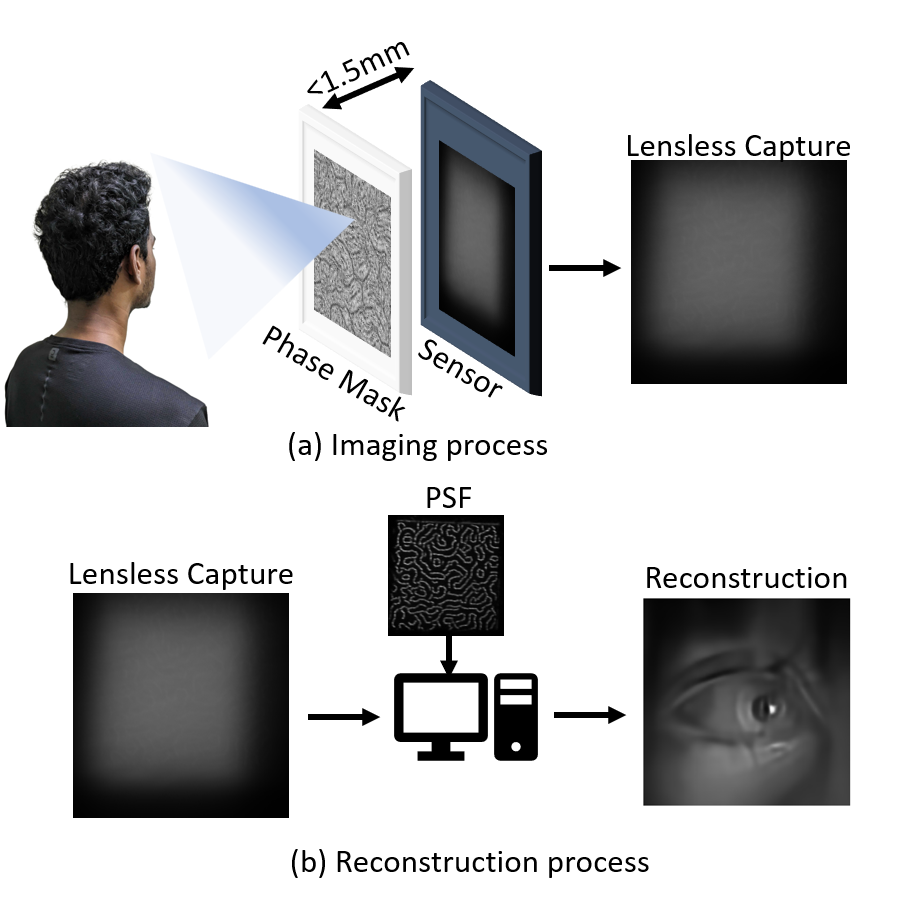}
  \caption{\textbf{PhlatCam Imaging Pipeline.} (a) Shows the forward imaging process and the lensless capture. Note that the capture doesn't resemble the scene it is imaging. (b) Reconstructing the scene back involves solving an inverse problem computationally.}
  \label{fig:phlatcam_fig}
\end{figure}

Due to the lack of a focusing element like a lens, a lensless camera capture is a multiplexed representation of the scene. To recover the original scene, one needs to computationally solve an inverse problem of estimating the scene $X$ given the measurement $Y$ and PSF $P$, which is challenging due to properties like large PSF support (which can be bigger than the scene projection $X$), finite sensor size, noise, etc. Although there exist different methods of lensless scene reconstruction in the literature, we will primarily focus on the two most popular methods -- FlatNet\cite{Khan_2020} and Wiener deconvolution. Please see Fig. \ref{fig:phlatcam_fig} for the lensless imaging and reconstruction process.

\section{FlatTrack:Lensless Gaze Estimation Dataset}
\begin{figure*}[t]
  \centering
  \includegraphics[width=\textwidth]{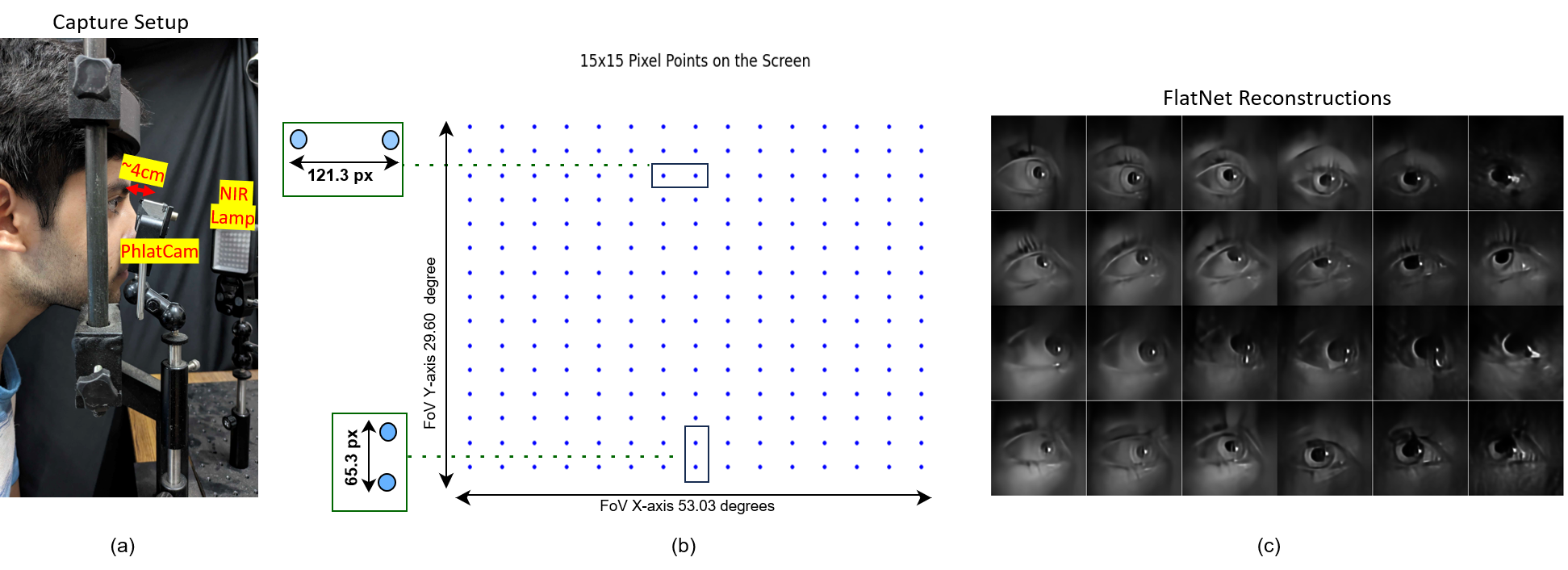}
  \caption{\textbf{FlatTrack Dataset.}(a) The capture setup used to collect the FlatTrack dataset. Note the small ($\sim 4cm$) distance between the camera and the eye. (b)The 15x15 grid on the monitor used as stimulus points to capture data has an FoV of 53.03 degree and 29.80 degree along x \& y axis. The pixel distance along x and y directions are 121.3 and 65.3 pixels respectively. (c) FlatNet reconstruction of 6 images each across the grid from 4 subjects are displayed}
  \label{fig:montage}
\end{figure*}
We use an NIR-PhlatCam designed for 700nm to capture the FlatTrack Gaze Dataset. We collected a comprehensive dataset comprising 20,475 lensless measurements meticulously gathered from 13 distinct subjects. The importance of user-specific data lies in the practical application of eye-tracking systems, where such systems are typically fine-tuned or calibrated for individual users using a small amount of personalized data. This calibration process is essential in real-world scenarios, such as AR/VR applications, where precise gaze estimation requires personalization.  

Each subject participated in a series of experiments in which they were tasked with focusing on stimuli presented on a computer screen. The stimuli moved along a 15x15 grid pattern as in Fig. \ref{fig:montage} within a 1080x1920 pixel area of the monitor\textit{ (DELL S2421HN)}, and their eye images were captured by the lensless camera. To ensure accuracy, subjects were seated 50 cm away from the screen, and before each round of data collection, calibration was performed to align the center of the eye with the origin on the screen. This process was repeated for each round of data collection for every subject. We also ensured that the distance between the eye of the subject and PhlatCam is approximately 4cm. We collected 4 to 5 rounds of data for each subject, each round comprising either 225 or 675 images (multiple measurements) corresponding to the 15 x 15 grid pattern. For each measurement, we also note the gaze point in pixels on the monitor and convert it to a unit vector in the gaze direction using the known monitor distance. 

The distance between two consecutive stimulus points along the x-direction in the grid pattern measures 121.3 pixels, while along the y-axis, it is 66.3 pixels. A margin of 50 pixels was also uniformly removed from all four sides of the grid to mitigate any potential edge effects. The field of view (FoV) subtended by the eye along the x-axis, spans 53.03 degrees. Similarly, the FoV along the y-axis extends to 29.6 degrees. Our capture setup, grid pattern, and lensless reconstruction of captured data are shown in Fig. \ref{fig:montage}.

Analysis of the angular differences between two consecutive grid points along the x-axis reveals a distinct trend characterized by an initial increase, peaking, and subsequent reduction. The minimum angle along the x-axis and y-axis is observed to be 3.21 degrees and 1.77 degrees, respectively. These trends are observed along the edges as the angular distance reduces with distance from the eye.

\section{Lensless Gaze Estimation Approach}
\begin{figure*}[tb]
  \centering
  \includegraphics[width=\textwidth]{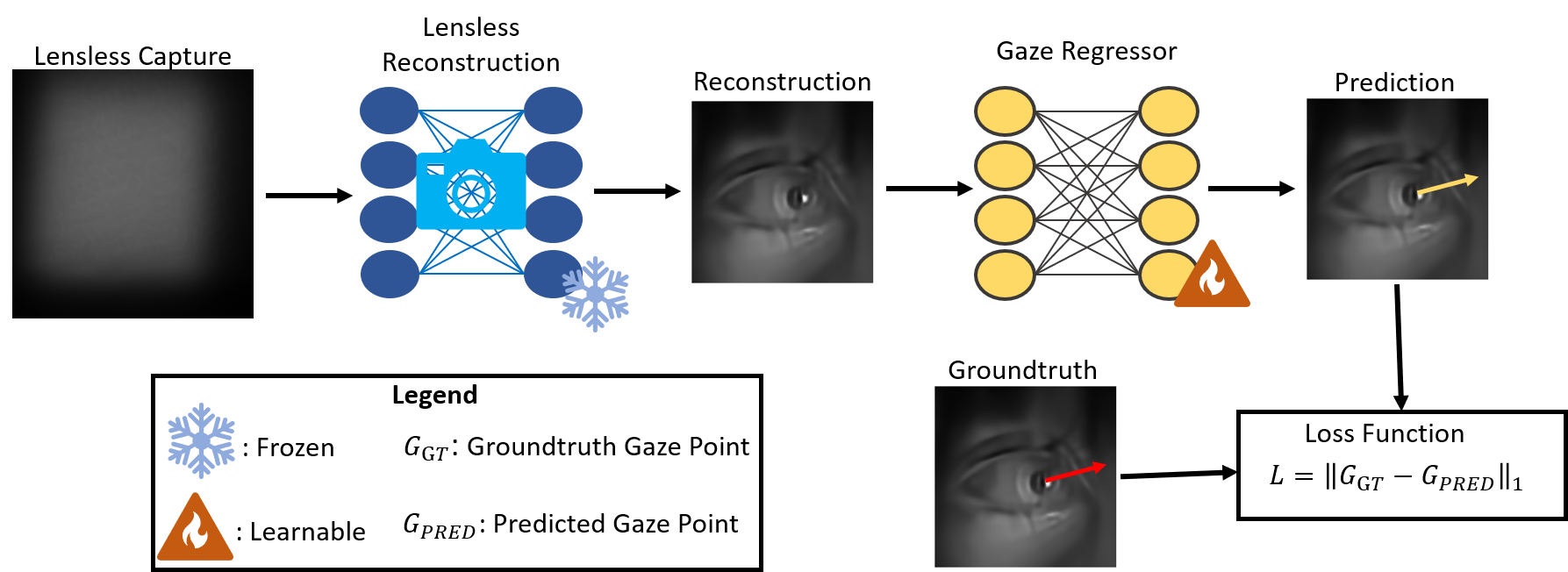}
  \caption{\textbf{Lensless Gaze Estimation Pipeline.} We follow a two-stage approach. In the first stage, we use a fixed lensless reconstruction algorithm to obtain the scene estimate. In the second stage, a gaze regression neural network predicts the gaze vector given the reconstruction. Using the loss function, the gaze regressor is updated while the reconstruction algorithm is frozen.}
  \label{fig:pipeline}
\end{figure*}
Since lensless measurements are globally multiplexed and do not resemble the scene, we use a two-stage approach: in the first stage, we use a reconstruction algorithm to obtain the scene estimate from the lensless captures; in the second stage, we use a gaze-regressor neural network to obtain the gaze direction. We discuss the approach in detail below.

\textbf{Scene Reconstruction.} Given the lensless forward model described in Eq. \ref{eq:lensless_eq}
Scene reconstruction from lensless capture involves solving an inverse problem of estimating the scene $X$, given the measurement $Y$ and PSF $P$, i.e., 
\begin{equation}\label{eq:reconstruction_eq}
    \hat{X} = \mathcal{F}(Y,P),
\end{equation}
where $\hat{X}$ is the scene estimate and $\mathcal{F}$ is the reconstruction algorithm. In this work, we experimented with two different approaches for $\mathcal{F}(.)$ -- Tikhonov-regularized least-squares (also known as Wiener deconvolution) and a learning-based algorithm called FlatNet\cite{Khan_2020}. Tikhonov Regularization involves solving the following equation, where Y is the lensless measurement, P is the PSF, and $X_T$ is the scene estimate. \begin{equation}
    \hat{X}_T = \arg \min_X \left\| Y - P * X \right\|_F^2 + \gamma \left\| X \right\|_F^2
\end{equation}. This equation has the following closed-form solution, where $F(.)$ is the Fourier transform and gamma is set to 1e-5. \begin{equation}
    \hat{X_T} = F^{-1} \left( \frac{(F(P))^* \odot F(Y)}{|F(P)|^2 + \gamma} \right)
\end{equation}For the learning-based FlatNet, we use pre-trained weights obtained by training it on a natural image dataset. FlatNet consists of two stages: (1) an inversion stage that maps the measurement into a space of intermediate reconstruction by learning parameters within the forward model formulation, and (2) a perceptual enhancement stage that improves the perceptual quality of this intermediate reconstruction. These stages are trained together in an end-to-end manner.

\textbf{Gaze Regressor.} Once the scene estimate $\hat{X}$ is obtained, we use a gaze regression network to finally predict the gaze direction. Our network first predicts the unit vector corresponding to the gaze direction. We then project this 3D gaze direction to a 2D gaze point (in pixels) on the monitor using the known distance between the monitor and the eye.

\textbf{Loss Function.} Once the gaze points on the monitor are estimated, we then use L1 loss on the predicted and groundtruth gaze points. More specifically, we use,
\begin{equation}\label{eq:loss_eq}
    \mathcal{L} = ||G_{GT} - G_{PRED}||_1,
\end{equation}
where $G_{GT}$ and $G_{PRED}$ are the ground-truth gaze point and predicted gaze point, respectively. Finally, we use this loss to update the gaze regression stage while keeping the reconstruction stage frozen. We show our gaze estimation pipeline in Fig. \ref{fig:pipeline}.

\section{Experiments}
In this section, we provide details of the implementation of our training strategy and a comprehensive evaluation of the gaze estimator. We benchmark various eye-tracking algorithms on our FlatTrack dataset and evaluate the performance on angular error and inference times. We also perform a quantitative comparison of lens-based v/s lensless imaging systems. We also assess the different lensless reconstruction methods: Wiener and FlatNet. Finally, we analyze the average per-pixel error of gaze estimation.


\subsection{Implementation Details}\label{impl_det}
The FlatNet model is pre-trained on measurements simulated using Eq. \ref{eq:lensless_eq} from the natural image MIRFLICKR dataset\cite{huiskes2008mir}. We freeze this pre-trained FlatNet model for the reconstruction of our captured measurements. 

We evaluate the performance of the following gaze estimation models: ResNet-18\cite{He_2016}, EyeCOD \cite{You_2022}, and MobileNetv2\cite{Sandler_2018}.
Of the 4 or 5 rounds of images captured for each subject, we keep one hold-out test round for each subject, which is later utilized to evaluate our models. We split the remaining dataset into training and validation sets (80:20). For each subject, we crop out the eye region from the reconstruction. We also perform random affine transformations on the reconstructed eye images to augment our data and to deal with small misalignments among the different sets.  

Our models predict a 3D unit gaze vector corresponding to a measurement. We follow a common training strategy for all models, which involves pre-training the models on the training dataset prepared above, followed by a subject-specific fine-tuning in which we freeze all the model parameters except the last 2 layers and fine-tune for the specific subject's training data. For training the models, 3D gaze vector outputs from the models are projected into 2D gaze positions using the known dimensions of the monitor and distance from the screen. Following this, L1 loss is calculated on these 2D labels and the 2D predictions for training the models. The common training regimen spans 50 epochs, employing the Adam optimizer with a weight decay of $0.0005$, an initial learning rate of $0.0001$, and a StepLR scheduler with a decay factor of 0.5 every 5 epochs. These hyperparameters remain consistent across all models for pre-training and fine-tuning. Finally, for each subject we test the fine-tuned model on the held-out set.

\subsection{Comparison of Lensless Reconstruction Methods}
In this section, we conduct a comparative analysis of the latency and performance of the ResNet-18 model using different lensless reconstruction methods, namely Weiner Deconvolution and FlatNet, on our captured dataset. Our primary focus is to evaluate the trade-off between accuracy and inference time.
The results of our comparison are summarized in Table~\ref{tab:weiner_vs_flatnet}.

\begin{figure}
 \centering
 \includegraphics[width=0.75\columnwidth]{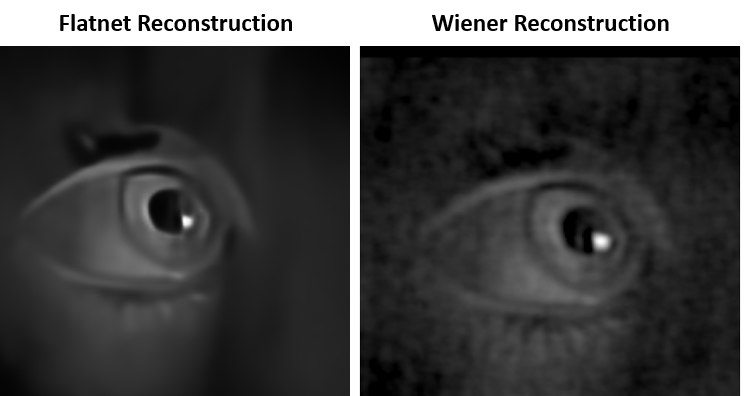}
 \caption{\textbf{Different Lensless Reconstruction Methods}. (a) FlatNet reconstruction. (b) Wiener deconvolution. FlatNet provides cleaner reconstruction.}
 \label{fig:modelplot}
\end{figure}

\begin{table}[h]
    \centering
    \begin{tabular}{|c|c|c|}
        \hline
        \textbf{Reconstruction}& \textbf{Average} & \textbf{Inference Time} \\
        \textbf{Method} & \textbf{Error (in deg)} & \textbf{(in ms)} \\
        \hline
        Weiner & 2.08$^\circ$ & \textbf{3.81} \\
        \hline
        FlatNet & \textbf{1.92$^\circ$} & 7.81 \\
        \hline
    \end{tabular}
    \caption{\textbf{Weiner v/s FlatNet.} We demonstrate that Flatnet based gaze-estimator achieves better accuracy while maintaining real-time gaze prediction with an equivalent to $\approx$130 fps or under \textbf{10ms}. The best result is in bold.}
    \label{tab:weiner_vs_flatnet}
\end{table}
Fig ~\ref{fig:modelplot} shows the example of a captured image reconstructed through both methods. The FlatNet reconstruction looks much cleaner and has less noise, which can be attributed to its better gaze estimation performance.

We observe minimal latency variation across both methods with a clear performance improvement in the case of FlatNet reconstruction; hence, we utilize the FlatNet method for our experiments. Note that these latency metrics are evaluated on an NVIDIA GeForce RTX3080 Ti graphics card and Intel Core i9-10900KF processor.

\subsection{Gaze Estimator Comparison} \label{methods}
In this section, we present the comparisons of the gaze estimation stage on the FlatTrack dataset. For this experiment, we fix the reconstruction stage as FlatNet and evaluate the performance of three different gaze estimators: ResNet-18\cite{He_2016}, EyeCOD \cite{You_2022}, and MobileNetv2\cite{Sandler_2018}.
We initialize ResNet-18 with ImageNet weights and replace the last layer to predict a 3D gaze direction.
EyeCOD is a gaze estimation model proposed by You et al. \cite{You_2022} for FlatCam captures. EyeCOD incorporates RITnet\cite{Chaudhary_2019} for semantic segmentation to crop out the Region of Interest(ROI) around the pupil consisting of the pupil, iris, and sclera, which is then passed on to ResNet-18 for gaze estimation. 
Since MobileNetv2 is specifically designed for on-the-edge devices and embedded vision applications where computational resources are limited, we also experiment with it.

We evaluate the performance of these gaze estimation architectures on the held-out test set for each subject based on the average angular error on the 3D gaze vectors and the inference times. We report two different errors in Table \ref{tab:resnet18}. The best-case error corresponds to the error on the held-out set for the subject with a minimum average error, while the average error is calculated across all the subjects. We also report the inference time evaluated on a single NVIDIA GeForce RTX3080 Ti graphics card.
\begin{table}[h!]
    \centering
    \begin{tabular}{|c|c|c|c|}
        \hline
        \textbf{Gaze} & \textbf{Best-Case} & \textbf{Average } & \textbf{Inference} \\
        \textbf{Estimator}&\textbf{ Error} & \textbf{Error} & \textbf{time (in ms)} \\
        \cline{1-1} \cline{2-4}
        \hline        
        ResNet-18 & \textbf{0.91$^\circ$} &  \underline{1.92$^\circ$} & \underline{7.81} \\
        \hline
        EyeCOD & \underline{0.95$^\circ$} & \textbf{1.82$^\circ$} & 22.72 \\
        \hline
        MobileNet & 1.08$^\circ$ & 2.43$^\circ$ &  \textbf{7.67} \\
        \hline
    \end{tabular}
    \caption{\textbf{Gaze Estimators Performance.} We use ResNet-18 as our base model for experiments as it has a comparable performance within \textbf{5\%} of the EyeCOD model with a \textbf{3X} reduction in inference time. The best result is in bold, while the second best is underlined.}
    \label{tab:resnet18}
\end{table}

We observe only $5\%$ improvement in performance using the EyeCOD methodology, although it bears a $3X$ increase in inference time. Clearly, Resnet-18 proves to be a better trade-off, so we perform further experiments using Resnet.


\subsection{Comparison of Lensed v/s Lensless Imaging Systems}
In this experiment, we assess the performance disparity between images captured by lensed and lensless cameras. As previously noted, there is no existing real-world dataset specifically for lensless eye tracking, which makes direct comparison with conventional datasets challenging. However, we have addressed this by simulating lensless measurements using the Davis-GS dataset\cite{Angelopoulos_2021}, which includes user-specific data. This is a crucial aspect, as our collected dataset also includes user-specific labels, making Davis-GS the most relevant benchmark for comparison. In contrast, datasets like OpenEDS2020\cite{palmero2020openeds2020} do not provide user-specific labels, which makes them less suited for evaluating fine-tuned eye-tracking systems.  More specifically, our evaluation focuses on comparing the performance of ResNet-18 on original lens images and lensless simulated images. The simulation process involves convolving the images with the point spread function (PSF) of our camera with some additive noise as described in Eq. \ref{eq:lensless_eq} and subsequently reconstructing them using FlatNet \cite{Khan_2020}. 

There are 27 subjects in the Davis-GS dataset, out of which we use 21 subjects' data for pre-training our gaze estimator and the remaining 3 subjects' data for fine-tuning and evaluation. Since the Davis-GS dataset consists of multiple images for random grid locations, we take one image for each grid pixel location (to avoid data imbalance) per subject and simulate lensless measurement. For fine-tuning the gaze estimator, we partition the 3 subject's data into train, validation, and test sets. After fine-tuning the model for respective subjects, we evaluate it on the test data for each subject. 


For pre-training and fine-tuning, we follow the same methodology as explained in Section \ref{impl_det}. However, here, due to the unavailability of 3D gaze vectors, we directly train the models on 2D gaze positions, which corresponded to the pixel location of the stimulus on the screen; hence, we replaced the last layer with a fully connected layer with the output size of 2 and borrow the other hyperparameters as mentioned in Section \ref{impl_det}. For fine-tuning, we unfreeze the last 2 layers of the pre-trained model and train on the test subject data. For the calculation of angular error, we first convert the predictions and labels to 3D gaze vectors using the known dimensions of the monitor and the distance of the subject from the screen. We report our findings on the test set using the average gaze angular error in degrees for the test set for each subject as summarized in Table~\ref{tab:lensed_vs_lensless}.

\begin{table}[h]
    \centering
    \begin{tabular}{|c|c|c|}
        \hline
        \textbf{Subject} & \textbf{Lensed} & \textbf{Lensless} \\
         & \textbf{(in deg)} & \textbf{(in deg)} \\
         \cline{1-1} \cline{2-2}\cline{3-3}
        \hline
        Subject25 & 1.79$^\circ$ & 1.84$^\circ$ \\
        \hline
        Subject26 & 1.72$^\circ$ & 1.81$^\circ$ \\
        \hline
        Subject27 & 1.67$^\circ$ & 1.62$^\circ$ \\
        \hline
    \end{tabular}
    \caption{\textbf{Lens v/s Lensless.} There is minimal difference in the performance for the lens and lensless images.}
    \label{tab:lensed_vs_lensless}
\end{table}
From Table \ref{tab:lensed_vs_lensless}, we can observe minimal difference in the performance of the two imaging systems, indicating that the lensless imaging system doesn't introduce a significant performance loss while allowing a compact form factor.

\subsection{ Lensless Gaze Estimation Analysis}
\begin{figure}[!ht]
  \centering
  \includegraphics[width=\columnwidth]{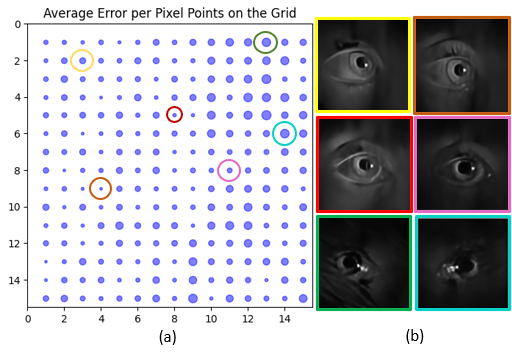}
  \caption{\textbf{Error Across the Grid}. The average per-pixel error for all 225 points in the grid across all subjects is displayed as a circle. Six grid points are circled, and their corresponding images from a subject dataset are displayed. The larger error corresponds to reconstructed lensless images of poor quality due to harsher illumination.}
  \label{fig:avgerror_plot}
\end{figure}
In this section, we analyze the errors observed in our experiments. To start with, the average error (across all subjects) at each grid pixel location is shown in Fig. ~\ref{fig:avgerror_plot}. The radius of the circle in the plot is proportional to the gaze prediction error. A circle with a higher radius implies more error. Six points are circled randomly from the grid, and their corresponding eye image reconstruction is also displayed on the side. 

We observe that the grid locations with lesser error correspond to better lensless reconstructed images, with pupils properly visible. Other points with high error are relatively dark images that lack pupil resolution. This is due to a lack of proper lighting as some of these points are farther away from the light source. A fix for these points with higher error would be to place additional light sources near them.

\section{Conclusion}
We propose a gaze-tracking framework for thin, lensless cameras. Lensless cameras, due to their ultra-thin form factor and lightweight nature, are attractive for applications requiring eye gaze tracking, such as Augmented/Virtual Reality. However, existing works on gaze estimations have not fully exploited this advantage. One of the bottlenecks for fully exploiting the benefits of lensless cameras for gaze tracking is the lack of a lensless gaze dataset. To deal with this, we propose the FlatTrack dataset - a first-of-its-kind NIR PhlatCam gaze dataset collected from 13 subjects and contains around 20K paired lensless measurements and gaze ground truth. Using a custom lensless gaze estimation pipeline evaluated on the FlatTrack dataset, we show that lensless gaze estimates can be of high fidelity. Furthermore, these gaze estimation algorithms can run at very high speeds ($>$125 fps) on a typical GPU. We hope the proposed dataset and the pipeline, when released, will drive more innovation in this promising space.

\textbf{Acknowledgments:} KM acknowledges funding support from IITM Pravartak Technologies Foundation, Department of Science and Technology, India (CRG/2023/007358) and Qualcomm Faculty Award 2024.

{\small
\bibliographystyle{ieee_fullname}
\bibliography{egbib}
}

\end{document}